\documentclass{emulateapj}

\slugcomment{Draft of \today}

\shorttitle{Evidence from SOFIA Imaging of Polycyclic Aromatic Hydrocarbon Formation along a Recent Outflow in NGC 7027}
\shortauthors{Lau et al.}

\usepackage{amsmath}
\usepackage{graphicx}
\usepackage[section] {placeins}
\usepackage{hyperref}
\usepackage{ulem}
\usepackage{color}

\newcommand{\beq}{\begin{equation}}
\newcommand{\eeq}{\end{equation}}

\begin{document}

\title{Evidence from SOFIA Imaging of Polycyclic Aromatic Hydrocarbon Formation along a Recent Outflow in NGC 7027}

\author{R. M. Lau\altaffilmark{1},
M. Werner\altaffilmark{1},
R. Sahai\altaffilmark{1},
M. E. Ressler\altaffilmark{1}
}
\altaffiltext{1}{Jet Propulsion Laboratory, California Institute of Technology, 4800 Oak Grove Drive, Pasadena, CA 91109, USA}

\begin{abstract}

We report spatially resolved (FWHM$\sim3.8-4.6''$) mid-IR imaging observations of the planetary nebula (PN) NGC 7027 taken with the 2.5-m telescope aboard the Stratospheric Observatory for Infrared Astronomy (SOFIA). Images of NGC 7027 were acquired at 6.3, 6.6, 11.1, 19.7, 24.2, 33.6, and 37.1 $\mu\mathrm{m}$ using the Faint Object Infrared Camera for the SOFIA Telescope (FORCAST).The observations reveal emission from Polycyclic Aromatic Hydrocarbon (PAH) and warm dust ($T_D\sim90$ K) from the illuminated inner edge of the molecular envelope surrounding the ionized gas and central star. The DustEM code was used to fit the spectral energy distribution of fluxes obtained by FORCAST and the archival infrared spectrum of NGC 7027 acquired by the Short Wavelength Spectrometer (SWS) on the Infrared Space Observatory (ISO). Best-fit dust models provide a total dust mass of $5.8^{+2.3}_{-2.6}\times10^{-3}$ $\mathrm{M}_\odot$, where carbonaceous large ($a=1.5$ $\mu$m) and very small ($a \sim12\AA$) grains, and PAHs ($3.1\AA<a<12\AA$) compose 96.5, 2.2, and 1.3 $\%$ of the dust by mass, respectively. The 37 $\mu$m optical depth map shows minima in the dust column density at regions in the envelope that are coincident with a previously identified collimated outflow from the central star.  The optical depth minima are also spatially coincident with enhancements in the 6.2 $\mu$m PAH feature, which is derived from the 6.3 and 6.6 $\mu$m maps. We interpret the spatial anti-correlation of the dust optical depth and PAH 6.2 $\mu$m feature strength and their alignment with the outflow from the central star as evidence of dust processing and rapid PAH formation via grain-grain collisions in the post-shock environment of the dense  ($n_H\sim10^5\,\mathrm{cm}^{-3}$) photo-dissociation region (PDR) and molecular envelope.

\end{abstract}

\maketitle

\section{Introduction}

Polycyclic aromatic hydrocarbon (PAH) molecules containing fewer than $\sim500$ carbon atoms are abundant and ubiquitous components of interstellar medium (ISM) characterized by their broad emission features at 3.3, 6.2, 7.7, 8.6, 11.3, and 12.7 $\mu$m (Tielens 2008 and ref. therein). Despite being smaller than $\sim10\AA$ in radius, observations of PAH features indicate that they account for $\sim10\%$ of carbon in the ISM (e.g. Allamandola et al. 1989, Draine \& Li 2007, Compi{\`e}gne et al. 2011). The details of PAH formation and destruction are, however, currently unclear. One of the most pervasive questions on interstellar PAHs is how they are replenished and/or survive destructive shocks driven by supernova explosions given that their theoretical destruction timescale is over an order of magnitude shorter than their production timescale from stars (e.g. Cherchneff, Barker, \& Tielens 1992; Micelotta et al. 2010a, b). Even though significant repopulation of PAH-sized grains is predicted to occur in $\sim100$ km $\mathrm{s}^{-1}$ shocks due to grain-grain collisions and the fragmentation of large graphitic grains (Jones et al. 1996), recent theoretical work shows that neither ``parent" nor ``daughter'' PAHs can survive the passage of shocks with velocities $\gtrsim100$ km $\mathrm{s}^{-1}$ (Micelotta et al. 2010a, b). This is in direct contrast with the detection of PAH features in shocked regions (e.g. Tappe al. 2006, Engelbracht et al. 2006, Armus et al. 2007). 

The circumstellar environment of carbon-rich post-Asymptotic Giant Branch (AGB) stars and young planetary nebulae (PNe) provide ideal laboratories to study the evolution of PAHs in the shocked regions impacted by collimated outflows from the central stars. Carbon-rich AGB stars, which are descendants of stars with an initial mass $<8$ $\mathrm{M}_\odot$, exhibit cool, dense outflows that are favorable environments for PAH formation (Latter 1991). These PAHs are believed to seed the growth of amorphous carbon and are therefore important components in the chemical pathways towards dust formation (Tielens 2008). Interestingly, PAH features are rarely observed towards carbon-rich AGB stars, which may be due to the cool effective temperatures of the AGB photosphere that are unable to excite the PAH vibrational and stretching modes. The PAH features are, however, present in spectra of carbon-rich post-AGBs and PNe where the circumstellar dust is heated by a radiation field blue from the hot degenerate core that is harder than the radiation field in the AGB-phase.

\begin{figure}[t]
	\centerline{\includegraphics[scale=0.35]{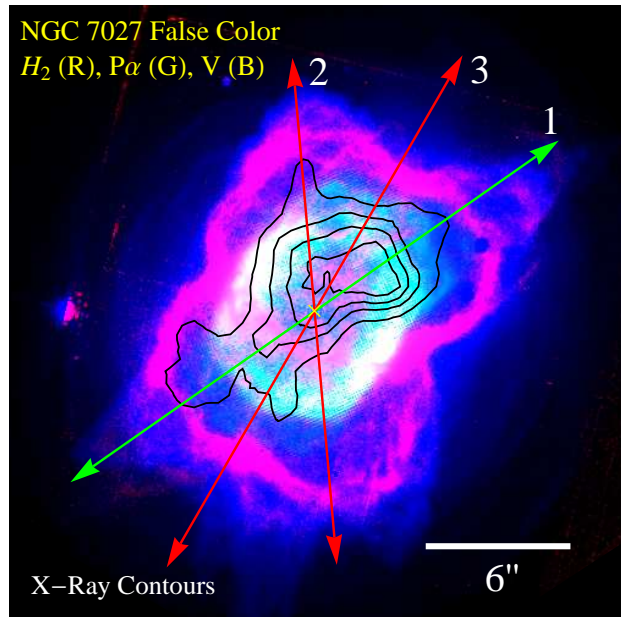}}
	\caption{False color image of NGC 7027 combining archival visible and near-IR observations with HST and x-ray observations with Chandra. Molecular hydrogen, Paschen-$\alpha$, and V-band emission are shown as red, green, and blue, respectively. X-ray contours are overlaid with levels corresponding to 20, 40, 60, and 80 \% of the peak x-ray flux. The bi-directional arrows labeled 1-3 correspond to the direction of the outflows identified by Cox et al. (2002). Outflow 1 (green) is believed to be the most recent and/or powerful outflow and shows a similar orientation to the bi-polar x-ray emission.}
	\label{fig:XRAYFCIm}
\end{figure}

NGC~7027 is a young and carbon-rich PNe, one of the brightest and best-studied of its breed (e.g. Gillett, Low, and Stein 1967; Becklin, Neugebauer, \& Wynn-Williams 1973; Moseley 1980). The compact $\sim10$ arcsec diameter central region is very bright at infrared and radio wavelengths, while more recent studies have identified an extended molecular shell, which is likely the envelope of the progenitor AGB star (Fig.~\ref{fig:XRAYFCIm}). Proper motion studies of features in the radio continuum in comparison with line of sight velocities show that the distance is $\sim1$ kpc and the dynamical age only $\sim1200$ yrs (Zijlstra et al. 2008). Infrared spectroscopy of the nebula reveals prominent PAH emission features as well as a high carbon abundance (Beintema et al. 1996, Bernard Salas 2001). Given the young age and relatively high mass ($\sim0.7$ $\mathrm{M}_\odot$, Zijlstra et al. 2008), it is not surprising that NGC~7027 is excited by a very hot white dwarf, which has a temperature in excess of 200,000 K and an estimated luminosity of 7700 $\mathrm{L}_\odot$ (Latter et al. 2000). 

Notably, the morphology of gas and dust in NGC~7027 exhibits significant deviations from spherical symmetry, which has been interpreted as evidence of interaction between the nebula and collimated high-velocity outflows from the central star (Graham et al. 1993, Kastner et al. 1994, Cox et al. 1997). Outflows from central stars are commonly observed in PNe and are believed to be one of the mechanisms responsible for shaping multi-polar morphologies (e.g. Sahai \& Trauger 1998; Sahai, Morris, \& Villar 2011). Kinematic observations of NGC~7027 indeed reveal asymmetric expansion velocities that suggest the passage of multiple collimated outflows through the ionized and molecular gas composing the nebula (Latter et al. 2000, Cox et al. 2002, L{\'o}pez et al. 2012). Three collimated outflows from the central star are identified by Cox et al. (2002) from kinematic information inferred from $H_2$ and Br-$\gamma$ radial velocities measurements. The outflows exhibit position angles of $-53^\circ$, $4^\circ$, and $-28^\circ$, which are referred to as outflows 1, 2, and 3, respectively (See Fig.~\ref{fig:XRAYFCIm}). Extended X-ray emission found along the axis of outflow 1 (Kastner et al. 2001) suggest that it is the most recent outflow ($<1500$ yr old). High red- and blue-shifted Br-$\gamma$ line velocities along outflow 1 of $\pm\,55$ km s$^{-1}$ with respect to the systemic velocity of the nebula ($v_\mathrm{LSR}\sim25$ km s$^{-1}$; Cox et al. 2002, Nakashima et al. 2010) reinforce this interpretation.

Bains et al. (2003) provide evidence of the interaction between outflow 1 and the nebula from their interpretation of a bright knot of red-shifted ionized gas at the north-west of the nebula, consistent with the axis of outflow 1. They find that the knot is bright due to higher temperatures as opposed to a density enhancement and claim the knot is associated with the receding, far-side of the nebula that is observable due to a ``breach" in the near-side from a high velocity outflow. Recent kinematic models of NGC~7027 from CO, $H_2$, and Br-$\gamma$ velocity measurements substantiate this hypothesis and find evidence of a ``hole" in the structure of all three emission components only along outflow 1 (Nakashima et al. 2010). The nebular structure along outflows 2 and 3 are only found to exhibit holes in the CO and $H_2$ components, which may alternatively be due to UV radiation and photodissociation as opposed to outflow interaction. This alternative mechanism for hole formation via photodissociation does not apply to Br-$\gamma$, which implies the hole along outflow 1 exhibits the strongest evidence of interaction with a high-velocity collimated outflow.

In this paper, we present mid-IR imaging of the young and carbon-rich PN NGC~7027 at wavelengths from 6 to 40 $\mu$m with $\sim4''$ resolution using the Faint Object Infrared Camera for the SOFIA Telescope (FORCAST). Compact PNe are ideal targets for SOFIA because of their high infrared surface brightness (e.g. Spuck et al. 2013, Werner et al. 2014). These are the first observations which extensively resolve the emission from the nebula at the wavelengths of its peak emission around 33.6 and 37.1 $\mu$m. The goal of our analysis is to study the morphology and energetics of the PAH and warm dust emission of NGC 7027 in the context of interactions with possible collimated outflows from the central star. In Sec. 2 we describe our observations with SOFIA and the data reduction. In Sec. 3, we report our results on comparing the mid-IR and 6.2 $\mu$m PAH feature morphology to archival HST imaging data, studying the color temperature and optical depth of the warm dust, and fitting dust models to our mid-IR spectral energy distribution. Lastly, in Sec. 4 we discuss the heating of dust in the nebula via trapped Lyman-$\alpha$ photons and report evidence of dust processing and rapid PAH formation along a recent outflow from the central star.

\section{NGC 7027 Observations}

\subsection{SOFIA/FORCAST Imaging Observations}

\begin{figure*}[t]
	\centerline{\includegraphics[scale=.48]{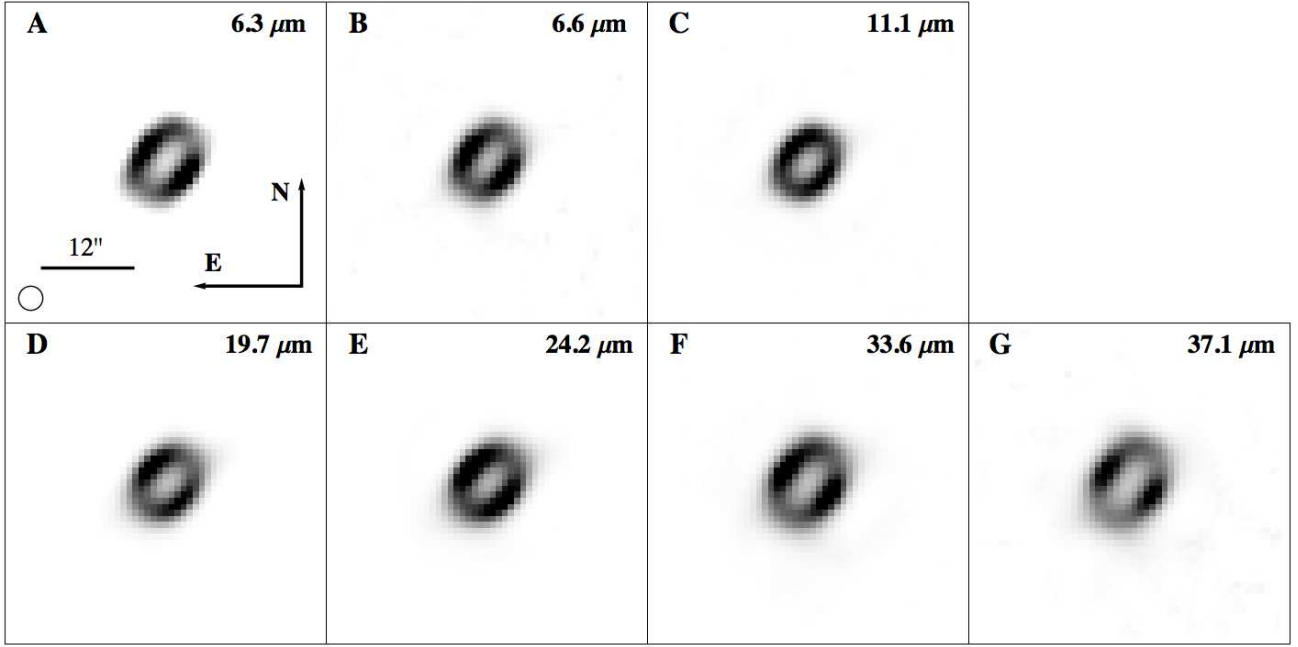}}
	\caption{De-convolved images of NGC 7027 at (A) 6.3 (B) 6.6 (C) 11.1 (D) 19.7 (E) 24.2 (F) 33.6 and (G) 37.1 $\mu$m. The de-convolved images have a uniform $3''$ FWHM Gaussian PSF. The FWHM of the deconvolved beam size is indicated on the lower left of (A).  North is up and east is to the left.}
	\label{fig:NebulaIR2}
\end{figure*}

Observations of NGC~7027 were made using FORCAST (Herter et al. 2013) on the 2.5-m telescope aboard SOFIA. FORCAST is a $256 \times 256$ pixel dual-channel, wide-field mid-infrared camera sensitive from $5 - 40~\mu\mathrm{m}$ with a plate scale of $0.768''$ per pixel and field of view of $3.4'\,\times\,3.2'$. The two channels consist of a short wavelength camera (SWC) operating at $5 - 25~\mu\mathrm{m}$ and a long wavelength camera (LWC) operating at $28 - 40~\mu\mathrm{m}$. An internal dichroic beam-splitter enables simultaneous observation from both long and short wavelength cameras. A series of bandpass filters are used to image at selected wavelengths.

SOFIA/FORCAST observations of NGC~7027 were taken on Basic Science Flight 59 on May 21, 2011 at 6.3, 6.6, 11.1, 19.7, 24.2, 33.6, and 37.1 $\mu\mathrm{m}$. The 6.3 and 6.6 $\mu$m narrow band filters have bandwidths of $\Delta\lambda=0.14$ and $0.24$ $\mu$m, respectively. The broadband filters 11.1, 19.7, 24.2, 33.6, and 37.1 have bandwidths of $\Delta\lambda=0.95,\,5.5,\,2.9,\,1.9$ and $3.3$ $\mu$m, respectively. The 6.6 and 11.1 $\mu$m data were taken without the dichroic, while the 24.2 $\mu$m data was taken simultaneously with that at 37.1 $\mu$m, and the 19.7 $\mu$m data simultaneously with that at 33.6 $\mu$m.  The total exposure time was 956 s at 6.3 $\mu$m, 972 s at 6.6 $\mu$m, 718 s at 11.1 $\mu$m, 841 s at 19.7 and 33.6 $\mu$m, 700 s at 24.2 $\mu$m, and 582 s at 37.1 $\mu$m. Chopping and nodding were performed to subtract fluctuations in the sky and telescope background. The chopping secondary on SOFIA was configured to chop with an amplitude of 85'' at 4 Hz and the telescope was nodded perpendicular to the chop direction with a 85'' throw. This chopping and nodding strategy made it possible to keep an image of the nebula within the field of view of the array continually during the observations. 

Calibration of the images was performed by observing standard stars and applying the resulting calibration factors as described in Herter et al. (2013). Raw data were processed applying the latest techniques for artifact removal and calibration (Herter et al. 2013). The 1-$\sigma$ uncertainty in calibration due to photometric error, variation in water vapor overburden, and airmass is $\pm7\%$; however, due to flat field variations ($\sim15\%$), which we are unable to correct for, we conservatively adopt a 1-$\sigma$ absolute photometric uncertainty of $20\%$. The integrated flux measurements of NGC 7027 across all wavelengths shown in Tab.~\ref{tab:Flux} have a signal-to-noise ratio greater than $\sim100$. 

A Richardson-Lucy deconvolution routine was performed on the processed NGC~7027 images in order to obtain higher spatial resolution and a uniform point spread function (PSF) at all wavebands. Due to the unstable PSF elongation in the cross-elevation direction in the earlier SOFIA flights, the standard stars observed during the flight were not used for the deconvolution. Instead, an artificially generated Gaussian PSF with a full-width at half maximum (FWHM) of the average 1-dimensional FWHM of the northeast and southwest edges of the NGC~7027 nebula was used for the deconvolution at each respective waveband. The resulting deconvolved images are shown in Figure~\ref{fig:NebulaIR2} with uniform $\sim3''$ resolution elements.\footnote{Level 3 processed data of NGC~7027 is also publicly available on the SOFIA Data Cycle System Science Archive (\url{https://dcs.sofia.usra.edu/}).}


\begin{deluxetable}{ccccccc}
\tablecaption{Observed Mid- and Far-Infrared Fluxes of NGC 7027 (in Jy)}
\tablewidth{240pt}
\tablehead{$F_{6.3}$ & $F_{6.6}$ &$F_{11.1}$ &$F_{19.7}$ &$F_{24.2}$ &$F_{33.3}$ &$F_{37.1}$  }

\startdata
	92 & 55 & 247 & 831 & 1549 & 1936 & 1838\\
\enddata

\tablecomments{Elliptical $13''\times16''$ apertures were used to extract the fluxes. The 1-$\sigma$ absolute flux calibration uncertainty is assumed to be $20\%$.}
	\label{tab:Flux}
\end{deluxetable}

\begin{figure*}[t!]
	\centerline{\includegraphics[scale=.32]{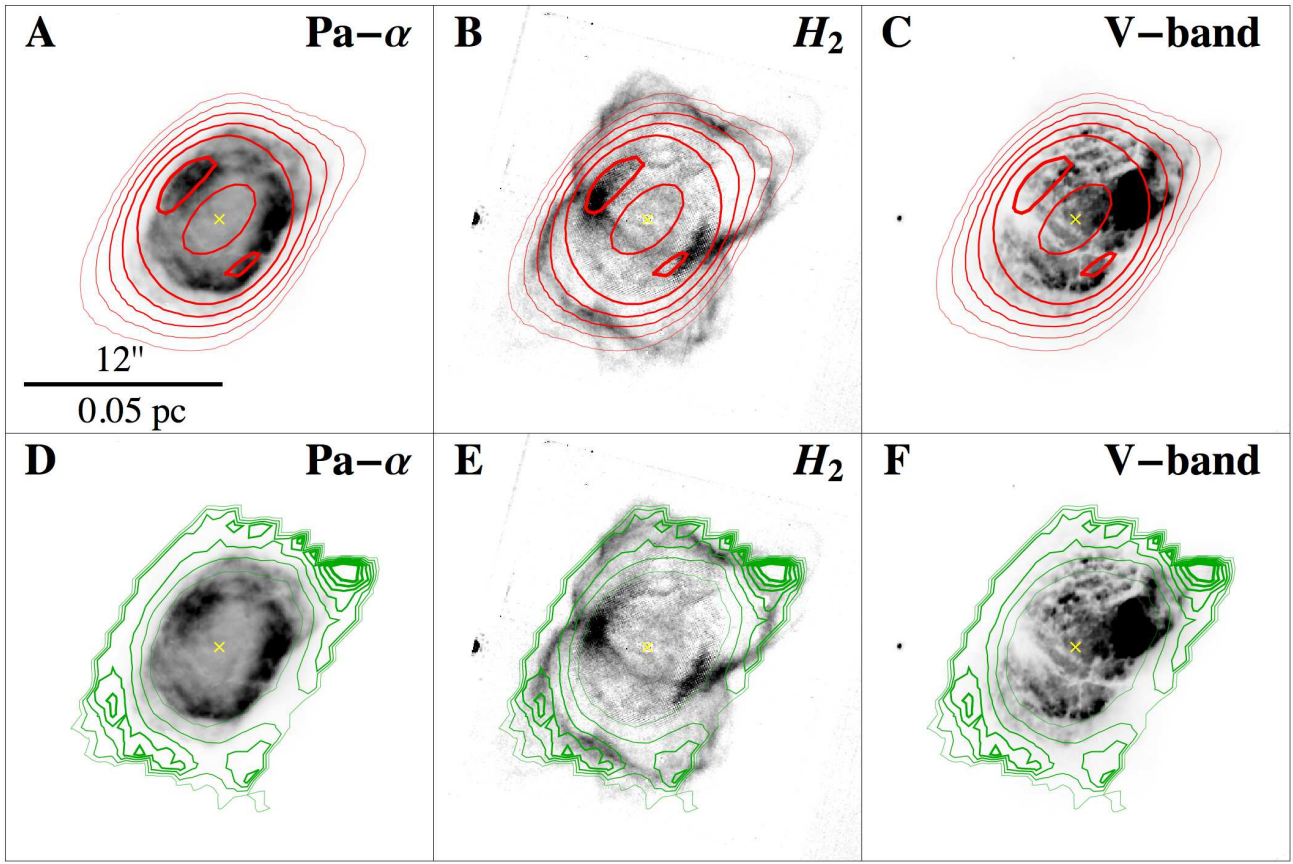}}
	\caption{On the top row are archival HST images of NGC 7027 tracing (A) Paschen-$\alpha$ (B) molecular hydrogen and (C) V-band. These images are overlaid with the de-convolved 19.7 $\mu$m emission contours with levels corresponding to 0.25, 0.5, 1.0, 2.0, 4.0 and 8.0 Jy/pix. 0.25 Jy/pix corresponds to 5 times the standard deviation in the background flux. On the bottom row are the same HST images as the top row overlaid with contours of $I_\mathrm{PAH6.2}$ with levels corresponding to 20, 30, 40, 50, 60, 70 and 80 $\%$ of the peak value of $I_\mathrm{PAH6.2}$ ($3.60\pm0.78$). The mid-IR imaging and HST maps are aligned within an error of $\sim1$ FORCAST pixel ($\sim0.75$''). North is up and east is to the left.}
	\label{fig:PNImall}
\end{figure*}

\subsubsection{$6.2$ $\mu$m PAH Emission Feature Strength}
Imaging with the 6.3 and 6.6 $\mu$m filters can be used as a diagnostic to approximate the equivalent width of the 6.2 $\mu$m PAH emission feature, $I_\mathrm{PAH6.2}$, since the 6.3 $\mu$m filter is broad enough to include the 6.2 $\mu$m PAH feature and the 6.6 $\mu$m filter traces continuum emission. The ratio of the differential flux between 6.3 and 6.6 $\mu$m and the 6.6 $\mu$m continuum flux is used to define $I_\mathrm{PAH6.2}$ as follows: $I_\mathrm{PAH6.2}=\frac{F_{6.3}-F_{6.6}}{F_{6.6}}$, where $F_\lambda$ is the flux at wavelength $\lambda$.

\subsection{Archival Visible, Near-IR, and X-Ray Imaging Data}

V-band (F555W) and near-IR  (F187N, F190N, and F212N) imaging data of NGC~7027 were obtained with the \textit{Hubble Space Telescope} (HST) on 1995 August 21 (PI: H. Bond, Proposal ID: 6119) and 1997 October 22 (PI: W. Latter, Proposal ID: 7365), respectively. The F555W filter on the Wide Field Planetary Camera 2 (WFPC2) has a mean wavelength, $\lambda_m$, of 5439 {\AA} and bandwidth,$\Delta\lambda$, of 1232 \AA. The F187N ($\lambda_m=18738$ {\AA}, $\Delta\lambda=192$ {\AA}), F190N ($\lambda_m=19005$ {\AA}, $\Delta\lambda=174$ {\AA}), and F212N ($\lambda_m=21213$ {\AA}, $\Delta\lambda=206$ {\AA}) images were taken with the Near Infrared Camera and Multi-Object Spectrometer (NICMOS) Camera 2. Maps of the Paschen-$\alpha$ ($\lambda =1.87 $ $\mu$m) and molecular hydrogen ($H_2$; $\lambda=2.12$ $\mu$m) line emission from NGC~7027 were produced from the F187N, F190N, and F212N images.

X-ray images of NGC 7027 were taken by the \textit{Chandra X-ray Observatory} with the Advanced CCD Imaging Spectrometer (AICS) on 2000 June 1 (PI: J. Kastner, Obs. ID: 588). As performed by Kastner et al. (2001), the image was convolved with Gaussian function with a FWHM of 2 pixels ($\sim1''$) to improve the signal-to-noise ratio.


\section{Results and Analysis}

\begin{figure*}[t]
	\centerline{\includegraphics[scale=.45]{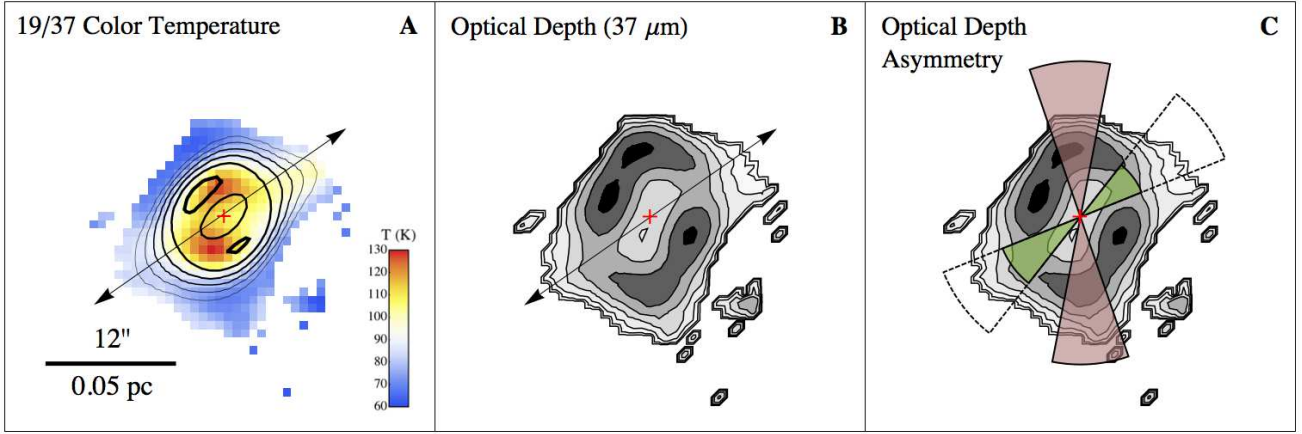}}
	\caption{(A) 19/37 color temperature map of NGC 7027 overlaid with an arrow indicating the direction of outflow 1 and 19.7 $\mu$m emission contours with levels corresponding to  to 0.25, 0.5, 1.0, 2.0, 4.0 and 8.0 Jy/pix. (B) Contour map of the 37 $\mu$m optical depth with contours corresponding to $\tau_{37}=$ 0.0012, 0.0024, 0.0048, 0.0096, 0.0192, 0.0384, and 0.0768 and overlaid with the same outflow 1 arrow as (A). (C) The optical depth map in (B) is overlaid with colored ``slices" aligned with outflow 1 (green) and 2 (red) centered on the central star whose length corresponds to the integrated optical depth in region subtended by the angular coverage of the slices. The dashed outline along the outflow 1 slice shows the size of the outflow 2 slice. Red crosses indicate the position of the central star. North is up and east is to the left.}
	\label{fig:PNCTOD}
\end{figure*}

\subsection{Dust, Gas, and PAH Morphology} 

NGC 7027 exhibits a nearly identical ellipsoidal morphology across all the observed mid-IR wavelengths (Fig.~\ref{fig:NebulaIR2}) with semi-major and semi-minor axes of $\sim4$ and $\sim3$'', respectively. A region of reduced brightness is seen at the center of the ellipse in each mid-IR image as well. The position angle of the main axis of symmetry is approximately $-30^\circ$, consistent with the morphology of the HII and photo-dissociation region (PDR; Latter et al. 2000, Cox et al. 2002). Emission across all mid-IR wavelengths, which primarily traces warm dust, is enhanced by a factor of $\sim2$ along the minor axis of the ellipse. In the de-convolved images there are faint, extended features that deviate from the symmetry of the ellipsoidal shell, which can be seen in the low 19.7 $\mu$m flux contours in Fig.~\ref{fig:PNImall}A - C. The asymmetric features extend $\sim8$'' from the center of the nebula to the NW and SE at an angle of $\sim-55^\circ$ and exhibit fluxes of $\sim10\%$ of the measured peak flux along the equatorial edge of the shell. 


Contours of the de-convolved 19.7 $\mu$m emission map and $I_\mathrm{PAH6.2}$ ($=\frac{F_{6.3}-F_{6.6}}{F_{6.6}}$) are overlaid on archival HST images of NGC 7027 in Fig.~\ref{fig:PNImall}A - C and Fig.~\ref{fig:PNImall}D - F, respectively, and show that the mid-IR emission and $I_\mathrm{PAH6.2}$ differ considerably. The $I_\mathrm{PAH6.2}$ map was derived from the 6.3 and 6.6 $\mu$m maps for pixels with a flux value greater than 3 times the standard deviation of the background flux. $I_\mathrm{PAH6.2}$ exhibits a peak value of $3.60\pm0.78$ at the northwest edge of the nebula.

The HST images in Fig.~\ref{fig:PNImall} map Paschen-$\alpha$, molecular hydrogen, and V-band emission which trace the inner ionized region of the nebula, the hot neutral shell, and scattered light from the central star, respectively (Latter et al. 2000)\footnote{A detailed discussion of the near-IR and visible morphology of NGC 7027 is presented in Latter et al. (2000) and Cox et al. (2002).}. The morphology of the mid-IR emission closely resembles Paschen-$\alpha$ (Fig.~\ref{fig:PNImall}A) and appears bounded by the $H_2$ shell (Fig.~\ref{fig:PNImall}B). This suggests the mid-IR emission is dominated by dust within the HII region at the inner edge of the extended molecular envelope (e.g. Bieging et al. 1991, Graham et al. 1993b). 

The overall morphology of the 6.2 $\mu$m PAH feature resembles that of an ellipsoidal shell with a position angle similar to that at the other wavelengths. However, it is larger in size than the nebula in the mid-IR and more consistent with the $H_2$ morphology tracing the PDR (Fig.~\ref{fig:PNImall}E). Peaks in the PAH feature strength are detected at the NW and SE regions of the nebula, consistent with the asymmetric and extended emission at the longer mid-IR wavelengths. Spherically symmetric shells of the progenitor AGB-envelope that are apparent in the V-band emission in Fig.~\ref{fig:XRAYFCIm}) are $\sim1000$ yr-old, which implies that the asymmetric mid-IR $I_\mathrm{PAH6.2}$ features were formed more recently. The PAH shell exhibits semi-minor and major radii of 4.6 and 8.5'', respectively, which suggests that the PAHs are associated with the extended ($\gtrsim$9''; Bieging et al. 1991; Graham et al. 1993b) molecular envelope formed during the AGB phase. This agrees with ISOCAM-CVF observations that show the 6.2 and 11.3 $\mu$m PAH emission features are more extended than the ionized gas and 9.8 $\mu$m continuum emission (Persi et al. 1999). At inner positions coincident with the Paschen-$\alpha$ shell, the PAH feature strength decreases to $\lesssim20\%$ of the peak value of $I_\mathrm{PAH6.2}$. 




The presence of the outflows from the central star (see Fig~\ref{fig:XRAYFCIm})  were determined from the kinematics and morphology of Brackett-$\gamma$ and $H_2$ radial velocity measurements by Cox et al. (2002). The V-band, near, and mid-IR images have symmetric axes consistent with outflow 3. The faint, asymmetric mid-IR emission and $I_\mathrm{PAH6.2}$ peaks at the NW and SE of the nebula are consistent with the orientation of outflow 1. Evidence for interactions between the nebula and outflow 1 are apparent in the extended emission at the NW and SE regions of the V-band image. The V-band image exhibits a bright knot $\sim4''$ NW of the central star consistent with the axis of outflow 1 and coincident with the suggested ``breach" in the nebula reported by Bains et al. (2003). No enhanced emission is detected at the expected ``counter-breach" position along outflow 1 SE of the central star, which may be due to an asymmetric outflow (e.g. Danehkar, Parker, \& Steffen 2016) and/or higher extinction along the line of sight to the SE (Kastner et al. 2001).


\subsection{Color Temperature and Optical Depth}

\begin{figure*}[t]
	\centerline{\includegraphics[scale=1.2]{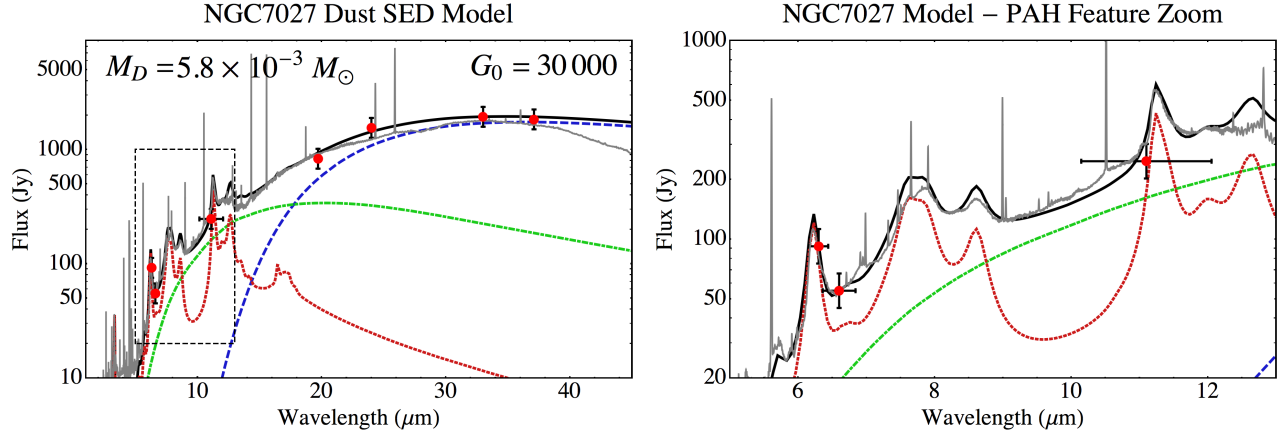}}
	\caption{(\textit{Left}) Best-fit DustEM model SED and (\textit{Right}) a zoomed-in plot of the PAH features (5 - 13 $\mu$m). The large grain (LG), very small grain (VSG), and PAH components in the plots are the blue dashed, green dot-dashed, and red dotted lines, respectively. The total sum of the model components is shown as the solid black line. FORCAST data are shown as bright red points with black error bars and the ISO/SWS spectrum is the gray solid line.}
	\label{fig:PNSED}
\end{figure*}

The color temperature map of NGC 7027 shown in Fig.~\ref{fig:PNCTOD}A is produced using the deconvolved (FWHM $\sim3''$) 19.7 and 37.1 $\mu$m images assuming the emission is optically thin and takes the form $F_\nu  \propto B_\nu  (T_d)\,  \nu^\beta$, where $B_\nu(T_d)$ is the Planck function which depends on the emission frequency, $\nu$, and dust temperature $T_d$, and $\beta$ is the index of the emissivity power-law. Pixel flux values less than a factor of 3 times the standard deviation of the background flux in each image were rejected from the map. The 19.7 and 37.1 $\mu$m maps were selected because they provide the longest wavelength baseline to derive temperatures, which excludes the wavebands that trace PAH emission features. No correction for interstellar or local extinction is applied since NGC 7027 is relatively nearby ($\sim910$ pc) and the relative reddening between 19.7 and 37.1 $\mu$m is less than the adopted $20\%$ photometric uncertainty assuming a mean visual extinction of $A_V\sim3$ (Robberto et al. 1993) within the nebula.

Since NGC 7027 is a carbon-rich nebula, it is assumed that dust is primarily composed of amorphous carbon, and an emissivity index of $\beta=1.2$ is adopted (Colangeli et al. 1995). The 19/37 color temperature map is shown in Fig.~\ref{fig:PNCTOD}A overlaid with the 19.7 $\mu$m emission contours. The 1-$\sigma$ uncertainty in the 19/37 color temperature is $\sim\pm6$ K. Temperatures exhibit peak values of $\sim130$ K at the inner northern and southern edges of the nebula, $\sim3$'' from the central star, and do not coincide with the regions of peak mid-IR emission. The NW and SE regions of the nebula that coincide with outflow 1 show temperatures of $\sim100$ and $\sim90$ K, respectively, which are hotter than the temperatures at locations equidistant from the central star ($\sim 60 - 70 $ K). A color temperature map derived with the 33.6 and 37.1 $\mu$m images, which are less affected by extinction, reveals a similar overall thermal structure (i.e. same positions of temperature maxima and minima) as the 19/37 map, validating the decision to ignore the differential extinction across the nebula. 

An optical depth map of the dust emission at 37.1 $\mu$m (Fig.~\ref{fig:PNCTOD}B) is derived from the color temperature and 37.1 $\mu$m emission maps by evaluating

\beq
\tau_{37\mu\mathrm{m}}=\frac{F_{37\mu\mathrm{m}}}{\Omega_p \,B_{37\mu\mathrm{m}}(T_D)}
\eeq

at each pixel position, where $\Omega_p$ is the solid angle subtended by a pixel ($1.4\times10^{-11}$ sr). The morphology of the optical depth map shown in figure~\ref{fig:PNCTOD}B is consistent with an ellipsoidal shell with a position angle identical to that of the mid-IR emission maps ($\sim-55^\circ$). Optical depths peak along lines of sight towards the equatorial and northeast regions of the nebula and exhibit maximum values of $\tau_{37\mu\mathrm{m}}=0.09\pm0.03$. The mean optical depth, $\langle{\tau}_{37\mu\mathrm{m}}\rangle\sim0.04$, implies a visual extinction of $A_V\sim4$ assuming the $R_V=3.1$ Weingartner \& Draine (2001) extinction curve where $A_{V}/A_{37}\approx93$.

At the NW and SE regions coincident with outflow 1, the optical depth map shows distinct minima that are $\sim25\%$ of the peak values along the equator. Figure~\ref{fig:PNCTOD}C illustrates the deviations from azimuthal symmetry in the integrated optical depth along outflows 1 and 2 ``slices'' subtended by identical angles. The general morphology of the optical depth resembles that of the molecular envelope of NGC 7027 as traced by the CO (1 - 0) emission at its systemic velocity (25 km $\mathrm{s}^{-1}$, Graham et al. 1993b; see Fig. 7 in Cox et al. 2002). Additionally, recent HCO$^+$ (J = 3-2) observations by Huang et al. (2010) taken with a similar spatial resolution ($3''$) to the FORCAST imaging data closely resemble the optical depth map and indicate the dust is primarily located in the PDR and inner edge of the molecular envelope. Note that the 37.1 $\mu$m optical depth map is sensitive to regions of warm dust emission and therefore does not trace cold ($T_d\sim20$ K; Sanchez Contreras et al. 1998) distributions of dust deeper into the extended molecular envelope.

\subsection{Dust Spectral Energy Distribution Model}

\begin{deluxetable*}{cccccc}
\tablecaption{Best-fit Dust SED Model Properties}
\tablewidth{500pt}
\tablehead{Dust Component & $M_d\,(\mathrm{M}_\odot)$ &$f_M\,(\%)$ &$a_\mathrm{min}\,(\mu\mathrm{m})$ &$a_\mathrm{max}\,(\mu\mathrm{m})$ &$a_0\,(\mu\mathrm{m})$ }

\startdata
	LG & $5.6^{+2.3}_{-2.4}\times10^{-3}$ & $96.6^{+0.9}_{-5.0}$ & - & - & $1.5^{+0.9}_{-0.2}$\\
	VSG & $1.1^{+0.8}_{-0.4}\times10^{-4}$& $1.9^{+3.4}_{-0.8}$ & $12\times10^{-4}$ & $20\times10^{-4}\lesssim a \lesssim0.4$ & -\\
	PAH & $8.9^{+1.6}_{-2.3}\times10^{-5}$ & $1.5^{+1.5}_{-0.7} $& $3.1\times10^{-4}$ & $12\times10^{-4}$ & $6.4\times10^{-4}$\\
\enddata

\tablecomments{The total dust mass from the best fit model is $5.8^{+2.3}_{-2.6}\times10^{-3}$ $\mathrm{M}_\odot$. $M_d$ is the dust mass of each dust component, $f_M$ is the percentage of the component's dust mass to the total dust mass, $a_\mathrm{min}$ and $a_\mathrm{max}$ are the minimum and maximum limits of the grain size in the component, and $a_0$ is the size of the single-sized LG distribution and the center of the log-normal PAH size distribution. The indicated 1-$\sigma$ errors of the model parameters are based on the reduced-$\chi^2$ from least squares fitting.}
	\label{tab:dust}
\end{deluxetable*}

Dust models are produced using DustEM (Compi{\`e}gne et al. 2011), which is capable of generating spectra of very small grains (VSGs, $10\AA<a<100\AA$) and polycyclic aromatic hydrocarbons (PAHs). Models are fit to the mid-IR flux measurements of the entire NGC 7027 nebula (Tab.~\ref{tab:Flux}) and archival ISO/SWS spectra where the emission is assumed to be optically thin. The radiation field heating the HII region, where the dust emission primarily originates (see Fig.~\ref{fig:PNImall}A), is complicated since the 200,000 K white dwarf primarily emits photons with energies greater than the ionization threshold for hydrogen. Self-consistent radiative transfer models of NGC 7027 (Volk \& Kwok 1997; Hasegawa, Volk, \& Kwok 1999) show that the radiation field from the central HII region is a combination of visible continuum photons from the central star and line photons from the ionized gas that correspond to the local interstellar UV field scaled up by $\chi\sim3\times10^4$ (the dominant dust heating mechanism is discussed in Sec.~\ref{sec:Lyman}). A standard radiation field of Mathis et al. (1983) scaled up by a factor of $3\times10^4$ is therefore adopted for the DustEM models.

Initially, only a single uniform distribution of amorphous carbon with radii $a>0.1$ $\mu$m (Sanchez Contreras et al. 1998) was adopted. However, a large grain (LG; $a>0.1$ $\mu$m) distribution under predicts the measured fluxes at $\lambda\lesssim 11$ $\mu$m by over an order magnitude. Dust emitting at $\lambda\lesssim 11$ $\mu$m likely trace a distribution of smaller, transiently-heated grains (e.g. Draine \& Li 2001). Smaller sized grains exhibit higher temperatures under the same heating conditions as larger grains due to lower heat capacities. Additionally, PAHs/VSGs are not typically heated in equilibrium with the incident radiation field given their small absorption cross-sections and therefore exhibit much greater temperatures when struck by individual photons. Given the presence of PAHs in the nebula (this work, Beintema et al. 1996, Bernard Salas 2001), PAH and VSG components were introduced in the models in addition to the LG component. A log-normal distribution of $3.1-12$ $\AA$ PAHs centered at $a_0=6.4$ $\AA$ with a width $\sigma=0.1$ (Compi{\`e}gne et al. 2011) and a power-law size distribution of amorphous carbon VSGs with $a_\mathrm{VSG}^\mathrm{min}=12$ $\AA$ and an index $\alpha=-3.5$ are adopted for the models. The free parameters for the SED models were the LG size, $a_{LG}$, the maximum VSG size, $a_\mathrm{VSG}^\mathrm{max}$, and the dust mass abundances for LGs, VSGs, and PAHs ($M_{LG}$, $M_{VSG}$, and $M_{PAH}$, respectively). Since emission from the VSG component will be dominated by the smaller grains in the distribution, $a_\mathrm{VSG}^\mathrm{max}$ is not well constrained. The adopted and fitted parameters are provided in Tab.~\ref{tab:dust}. 

The best-fit SED model is shown in Fig.~\ref{fig:PNSED} (\textit{left panel}) overlaid with the spectrum from ISO/SWS and indicates that dust in the nebula by mass is primarily composed of the LGs, which we have assumed to be $=1.5$ $\mu$m-sized amorphous carbon grains. The total dust mass is $5.8^{+2.3}_{-2.6}\times10^{-3}$ $\mathrm{M}_\odot$. Figure~\ref{fig:PNSED} (\textit{right panel}) shows a zoomed-in plot of the PAH features between 5 - 13 $\mu$m. VSGs and PAHs are shown to contribute $\sim2\%$ and $\sim1.5\%$ to the total dust mass, respectively. Assuming a gas-to-dust mass ratio of 150, the results from the fit imply a total gas mass of $\sim0.9$ $\mathrm{M}_\odot$, an order of magnitude greater than the ionized gas mass ($\sim0.02$ $\mathrm{M}_\odot$, Latter et al. 2000) and $30\%$ less than the observed molecular gas mass ($\sim1.3$ $\mathrm{M}_\odot$; Santander-Garcia et al. 2012). The deviations from the dust-derived mass and the ionized and molecular gas masses are expected since the warm dust emission traces the HII region and PDR of the nebula, but does not trace colder dust in the dense, molecular envelope.

\section{Discussion}

\subsection{Dust Heating by Trapped Lyman-$\alpha$ Photons}
\label{sec:Lyman}

The central star of NGC 7027 exhibits a temperature and luminosity of $T_{eff}=200,000$ K and $L=7.7\times10^3$ $\mathrm{L}_\odot$, respectively (Latter et al. 2000). Based on our best-fit dust models, the total IR luminosity re-radiated by the warm dust is $\sim5.6\times10^3$ $\mathrm{L}_\odot$, which is $\sim70\%$ of the central star luminosity. The high effective temperature of the central star, however, implies that most of its luminosity ($\sim99\%$ assuming a $T_{eff}=200,000$ K blackbody) is in the form of ionizing photons ($E>10.2$ eV). Due to this luminosity discrepancy and the consistent morphology of the mid-IR and ionized gas emission, we claim that trapped Lyman-$\alpha$ photons that are rapidly absorbed and re-emitted in the HII region are responsible for heating the dust.

The Lyman-$\alpha$ luminosity, $L_{\mathrm{Ly}\alpha}$, can be estimated by

\beq
L_{\mathrm{Ly}\alpha}=V\,n_e^2(\alpha_B-\alpha_{2s1})h\nu_{\mathrm{Ly}\alpha}\,4\pi d^2,
\label{eq:TLy}
\eeq

where $V$ is the volume of the emitting HII region, $n_e$ is the electron density, $\alpha_B$ is the type-B recombination coefficient, $\alpha_{2s1}$ is the recombination coefficient for the two-photon process ($\alpha_{2s1}<\alpha_B$; Osterbrock \& Ferland 2006), and $d$ is the distance to NGC 7027 (910 pc). Radio flux measurements of free-free emission from the HII region, $S_{\nu,ff}$, can therefore be used to estimate $L_{\mathrm{Ly}\alpha}$ given that $S_{\nu,ff}=j_{\nu,ff}(n_e,T_e)\,V$, where $j_{\nu,ff}$ is the volume emissivity for free-free emission. Zijlstra et al. (2008) provide 15 GHz flux measurements of 5.8 Jy for NGC 7027. Assuming an electron temperature of $T_e=15,000$ K (Roelfsema et al. 1991, Latter et al. 2000), we estimate a Lyman-$\alpha$ luminosity of $L_{\mathrm{Ly}\alpha}\approx4\times10^3$ $\mathrm{L}_\odot$. The slight discrepancy between the Lyman-$\alpha$ and IR luminosity is likely due to dust heating from Lyman-$\alpha$ (and other line photons) ``leaking" from the HII region into the PDR (e.g. Volk \& Kwok 1997). Our results therefore indicate that dust is heated by trapped Lyman-$\alpha$ photons. 



\subsection{Evidence for Dust Processing and Rapid PAH Formation Along Outflow 1}

\begin{figure*}[t]
	\centerline{\includegraphics[scale=0.4]{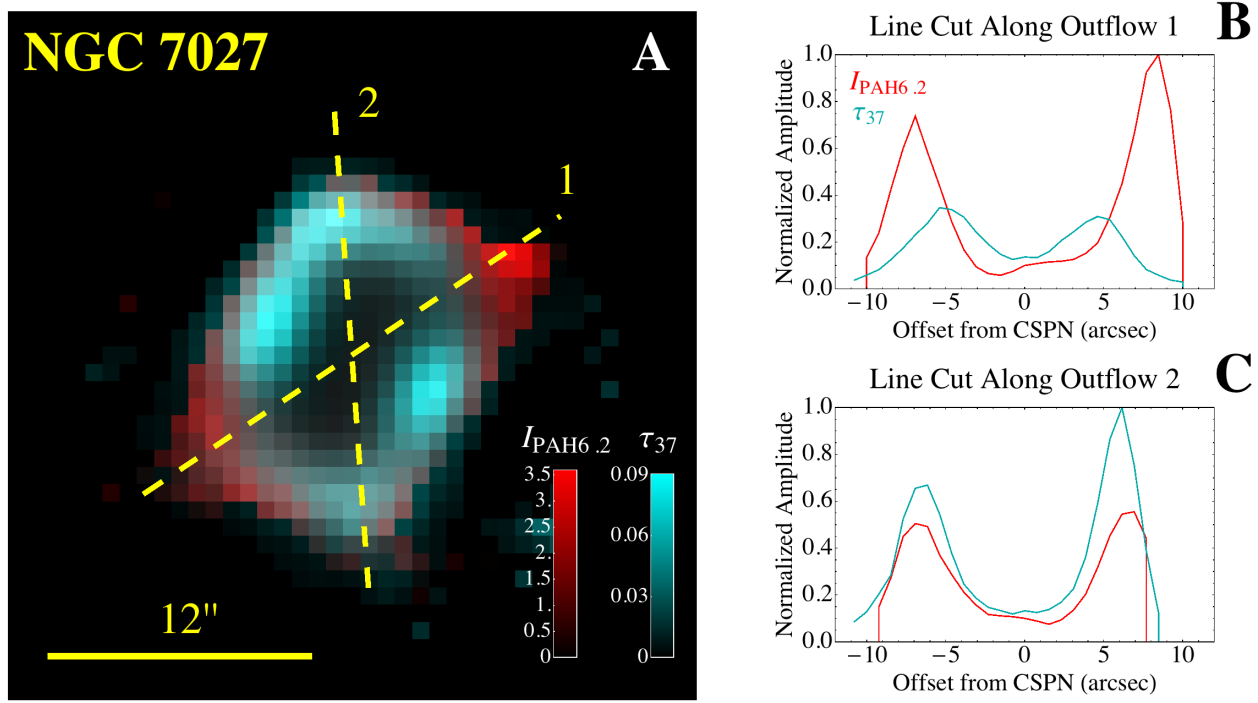}}
	\caption{(A) False color 37 $\mu$m optical depth (cyan) and $I_\mathrm{PAH6.2}$ (red) map of NGC~7027. Yellow, dashed lines indicate positions corresponding to outflow (B) 1 and (C) 2, where line cut measurements were made of the normalized amplitudes of $I_\mathrm{PAH6.2}$ (red) and $\tau_{37\,\mu\mathrm{m}}$ (cyan).}
	\label{fig:PNLCIm}
\end{figure*}



The optical depth minima in the nebula coincident with outflow 1 from the central star (Fig.~\ref{fig:PNCTOD}B) and deviations from azimuthal symmetry in the optical depth (Fig.~\ref{fig:PNCTOD}C) indicate that significant quantities of the LG dust distribution have been removed via sputtering and/or shattering. A comparison of line cut measurements of the optical depth and $I_\mathrm{PAH6.2}$ along outflows 1 and 2 shown in Fig.~\ref{fig:PNLCIm}A-C indeed reveal a noticeable spatial anti-correlation between the optical depth and $I_\mathrm{PAH6.2}$ along outflow. The observed 6.2 $\mu$m PAH emission feature, $I_\mathrm{PAH6.2}$, scales with both PAH abundance and the hardness of the incident radiation field. We claim that deviations in the strength of $I_\mathrm{PAH6.2}$ are mainly due to changes in PAH abundance and not variations in the hardness of the radiation field since PAHs throughout the nebula share a common heating source.  


The coincidence of the PAH emission peaks with the optical depth minima and outflow 1 suggests they are physically linked. Figure~\ref{fig:PNLCIm}A shows the spatial anti-correlation between $I_\mathrm{PAH6.2}$ and optical depth minima. Note that $\tau_{37\mu\mathrm{m}}$ is not sensitive to the optical depth of PAHs since $\tau_{37\mu\mathrm{m}}$ is derived from the 19.7 and 37.1 $\mu$m maps that trace emission from the LG distribution. Given the morphological evidence, we claim that the LGs were processed/destroyed and that PAHs were rapidly formed in the interaction between outflow 1 and the nebula.


Dust in shocked regions undergo sputtering due to collisions with ionized particles impacting the grain with kinetic energy higher than the surface binding energy of the grain. Sputtering can occur in two different modes that are distinguished by whether the collisions are isotropic with low relative velocity between the dust and gas (thermal) or if the collisions are due to the relative motion between the dust and gas (non-thermal). Large grains are also shattered in collisions with smaller grains with relative velocities $1\lesssim v\lesssim75$ km $\mathrm{s}^{-1}$ for graphitic/amorphous carbon dust (Jones et al. 1996). This shattering results in the redistribution of dust mass from large to smaller grain sizes, and is therefore predicted to be an efficient mechanism of PAH formation. In this section, we discuss the effects of dust destruction and shattering in the PDR due the shock driven by outflow 1. We show that the optical depth minima of the LGs are likely due to the destruction of dust via non-thermal sputtering and that the enhanced PAH abundance is due to rapid PAH formation from grain-grain collisions in the post-shock environment of the PDR and molecular envelope. 


\subsubsection{Dust Destruction}

We first consider the destruction of the large dust grains ($a\sim1$ $\mu$m) from thermal sputtering by collisions with ionized particles. The thermal sputtering timescale, $\tau_\mathrm{sp}$, must be comparable to or shorter than the radiative cooling timescale of the shocked gas, $\tau_\mathrm{cool}$ if significant thermal sputtering is expected to occur.

The electron temperature and density of the shock driven by outflow 1 into the PDR and molecular envelope, where the emitting mid-IR dust is located, can be estimated by the observed properties of the diffuse, shocked gas in the cavity of the nebula. X-ray observations of the bi-polar shocked gas that are believed to be associated with outflow 1 (see Fig.~\ref{fig:XRAYFCIm}) reveal $T^\mathrm{jet}_e\sim8\times10^6$ K and $n^\mathrm{jet}_e\sim150$ $\mathrm{cm}^{-3}$ (Kastner et al. 2001; Maness et al. 2003). Since the densities are greater in the PDR and molecular envelope ($n_e^\mathrm{PDR}\sim10^5$ $\mathrm{cm}^{-3}$), the strength of the shock in the higher density PDR will be attenuated by the square root of the ratio of the outflow and PDR densities (i.e. $T_e^\mathrm{PDR}=T_e^\mathrm{jet}\sqrt{n^\mathrm{jet}_e/n^\mathrm{PDR}_e}$). Assuming a solar abundance for gas composing the nebula, the radiative cooling timescale in the PDR and molecular envelope can be approximated by (Draine 2011)

\beq
\tau_\mathrm{cool}\approx0.14\,\mathrm{yr}\left(\frac{T^\mathrm{PDR}_e}{3\times10^5\,\mathrm{K}}\right)^{1.7}\left(\frac{n^\mathrm{PDR}_e}{10^5\,\mathrm{cm}^{-3}}\right)^{-1},
\label{eq:cool2}
\eeq

which is much faster than the cooling timescale of the diffuse, x-ray emitting gas interior to the envelope ($\sim2.5\times10^4$ yr).





Thermal sputtering is only important for shocks with $T_e\gtrsim10^6$ K (Tielens et al. 1994), therefore we do not expect short thermal sputtering timescales. The thermal sputtering timescale for graphitic grains (Draine \& Salpeter 1979) for a $T_e\sim3\times10^5$ K shock is approximately

\beq
\tau_\mathrm{sp}^\mathrm{th}\approx 1000\,\mathrm{yr}\left(\frac{a}{1\,\mu\mathrm{m}}\right)\left(\frac{n_H}{10^5\,\mathrm{cm}^{-3}}\right)^{-1},
\label{eq:sput}
\eeq

where $n_H$ is the hydrogen number density of the shocked medium. Since the cooling timescale in the PDR is much shorter than the thermal sputtering timescale for the shock driven by outflow 1, it is unlikely that the destruction of the LGs is due to thermal sputtering.

Next, we consider the effects of non-thermal sputtering. As the shocked PDR rapidly cools, the gas and the embedded magnetic field are both compressed. Due to their inertia, the LGs in the post-shock regions of the PDR will exhibit a differential velocity from the gas of $\sim3/4$ the shock speed. Since the grains are likely charged, they will undergo ``betatron acceleration" and gyrate around the magnetic field lines embedded in the compressed gas (McKee et al. 1987, Jones et al. 1994, 1996). Grains that are large enough such that the deceleration time due to gas drag ($\tau_\mathrm{drag}\propto a$) is longer than the cooling time of the shocked gas will be accelerated by the magnetic field and destroyed via non-thermal sputtering as the grain spirals around the field lines relative to the ``stationary" gas. McKee et al. (1987) determine the critical grain size, where the drag time equals the cooling time, to be

\begin{multline}
a_\mathrm{crit}=0.044\,\mu\mathrm{m}\left(\frac{v_s}{80\,\mathrm{km}\,\mathrm{s}^{-1}}\right)^4 \\ \left(\frac{\rho_\mathrm{gr}}{2.2\,\mathrm{g}\,\mathrm{cm}^{-3}}\right)^{-1}(1+1.52\, x_p \,\phi^2),
\label{eq:acrit}
\end{multline}

where $v_s$ is the shock velocity, $\rho_\mathrm{gr}$ is the bulk density of the dust grain, $x_p$ is the hydrogen ionization faction, and $\phi$ is the grain charge parameter. We note that the estimated velocity of the $3\times10^5$ K shock ($v_s\sim80$ km s$^{-1}$) is similar to the de-projected velocity of the PDR along outflow 1 ($\sim70$ km s$^{-1}$) as approximated by adopting an inclination of $53^\circ$, consistent with the position angle, and the $\pm55$ km s$^{-1}$ red- and blue-shifted Br-$\gamma$ line emission measured by Cox et al. (2002). Calculations of the grain charge parameter in $T_e\sim10^5$ K shocks indicate $|\phi|\lesssim3$ (e.g. McKee et al. 1987; Bocchio, Jones, \& Slavin 2014). For a fully ionized plasma ($x_p=1$) and adopting $|\phi|\approx3$, we find that the LGs inferred from our model ($a\sim1.5$ $\mu$m) will be greater than $a_\mathrm{crit}$ and therefore dominated by non-thermal sputtering in the shock driven by outflow 1. 

The rate at which a dust grain decreases in size due to traveling at a speed $v$ relative to ions $i$ is given by

\beq
\left(\frac{da}{dt}\right)=\frac{m_\mathrm{sp}}{2\rho_\mathrm{gr}}v\,n_H\sum A_i\,Y_i (E^\mathrm{kin}_i),
\eeq

where $m_\mathrm{sp}$ is the mass of the sputtered atom ($12\,m_p$ for amorphous carbon), $A_i$ is the abundance of ion $i$ relative to hydrogen, and $Y_i$ is the sputtering yield for ion $i$ which is a function of $E^\mathrm{kin}_i=0.5m_i\,v^2$. Adopting nebular abundances from Bernard Salas et al. (2001) and the Tielens et al. (1994) yields for hydrogen and helium ions, which dominate the non-thermal sputtering, and assuming $v\approx3/4 v_s=60$ km $\mathrm{s}^{-1}$, we determine a lifetime of $\sim675$ yr for 1.5 $\mu$m sized grains. Since the derived $\sim675$ yr non-thermal sputtering lifetime is shorter than the $\sim1000$ yr dynamical timescale of the nebula, we claim that significant dust destruction via non-thermal sputtering has occurred in regions of the PDR impacted by outflow 1. The LG lifetime is likely shortened when considering the effect of LG shattering via grain-grain collisions discussed in the following subsection.

\subsubsection{Dust Shattering and PAH Formation}

Grain-grain collisions in shocks will greatly impact the redistribution of mass from larger to smaller grain sizes as large grains are fractured into many smaller fragments (Jones et al. 1996). Grain collisions are not an efficient mechanism for grain destruction since vaporization requires collisions between large grains, which are rare due to their low number density.  Although VSGs and PAHs do not compose a large percentage of the total dust mass, they have a much higher number density since $n_d\propto a^{-3.5}$ and will therefore collide frequently with the large $\sim1$ $\mu$m sized grains. Importantly, the small grain distributions are weakly affected by betatron acceleration since $a_\mathrm{VSG/PAH}<a_\mathrm{crit}$ (Eq.~\ref{eq:acrit}), which implies a high relative velocity difference between small grains and the large grains that are gyrating around the compressed magnetic field lines. 

The collision timescale between large grains of size $a_\mathrm{LG}$ moving with velocity $v_\mathrm{LG}$ relative to small grains of number density $n_\mathrm{VSG}$ can be estimated by 

\beq
\tau_\mathrm{coll}\sim\frac{1}{\pi\,a_\mathrm{LG}^2\,n_\mathrm{VSG}\,v_\mathrm{LG}},
\label{eq:coll}
\eeq

and we can approximate the VSG number density as

\beq
n_\mathrm{VSG}\approx \frac{m_H\,n_H\,\chi_\mathrm{D/G}\,\chi_\mathrm{VSG/LG}}{4/3\pi\,a_\mathrm{VSG}^3\,\rho_\mathrm{gr}},
\eeq

where $\chi_\mathrm{D/G}$ and $\chi_\mathrm{VSG/LG}$ are the dust-to-gas and VSG-to-LG mass ratios. Adopting $\chi_\mathrm{D/G}=1/150$ and $\chi_\mathrm{VSG/LG}=0.02$, we evaluate the collision timescale between 10 $\AA$ and 1.5 $\mu$m grains to be

\begin{multline}
\tau_\mathrm{coll}\sim 930\,\mathrm{s} \left(\frac{a_\mathrm{LG}}{1.5\times10^{-4}\,\mathrm{cm}}\right)^{-2} \left(\frac{a_\mathrm{VSG}}{10\times10^{-8}\,\mathrm{cm}}\right)^{3} \\\left(\frac{n_H}{10^5\,\mathrm{cm}^{-3}}\right)^{-1} \left(\frac{v_\mathrm{LG}}{60\,\mathrm{km}\,\mathrm{s}^{-1}}\right)^{-1}\left(\frac{\chi_\mathrm{D/G}}{0.007}\right)^{-1}\\ \left(\frac{\chi_\mathrm{VSG/LG}}{0.02}\right)^{-1} \left(\frac{\rho_\mathrm{gr}}{2.2\,\mathrm{g}\,\mathrm{cm}^{-3}}\right).
\label{eq:tcoll}
\end{multline}

The grain-grain collision timescale is shorter than the non-thermal sputtering timescale for the LGs and the thermal sputtering timescale for VSGs and PAHs (Eq.~\ref{eq:sput}), implying that the small grain projectiles will survive through the cooling of the shocked gas and that numerous VSG/PAH-LG collisions will occur before the LGs are sputtered away. The overall outcome of the outflow-driven shock into the PDR is therefore a net production of VSGs and PAHs, which is apparent from the enhancement of the 6.2 $\mu$m PAH emission feature we observe along outflow 1 (Fig.~\ref{fig:PNLCIm}A and B). We note that Eq.~\ref{eq:tcoll} also shows that collisions between LGs will occur on timescales of $\sim2000$ yr, a factor of two longer than the dynamical age of the nebula. Hotter dust temperatures localized along outflow 1 (Fig.~\ref{fig:PNCTOD}) at the optical depth minima support the dust shattering/VSG production scenario since VSGs exhibit higher temperatures than large grains in identical heating conditions. This is due to the lower heat capacity of VSGs, which proportional to the total number of atoms composing the grain.


\subsubsection{Interactions Along Outflows 2 and 3?}

The maps of the 37 $\mu$m optical depth and strength of $I_\mathrm{PAH6.2}$ exhibit significant anti-correlation only along outflow 1 (see Fig.~\ref{fig:PNLCIm}). We therefore do not find evidence of PAH formation along outflows 2 and 3 in the PDR of the nebula. We consider the following explanations to account for this observation: 1) outflows 2 and 3 were weaker than outflow 1 and unable to breach the nebular shell, 2) outflows 2 and 3 are not associated with a true outflow but rather a bipolar UV-radiation field (e.g. Nakashima et al. 2010), and/or 3) outflows 2 and 3 are significantly older than outflow 1 and the VSGs/PAHs formed in the interaction with the nebula have propagated far enough away from the central star such that they are not detected. 

The absence of optical depth minima coincident with the orientation of outflows 2 and 3 suggests that the envelope has not been breached recently by these outflows. We estimate the timescale for a breached envelope to ``equilibrate" to a more uniform density to be an orbital period around the central star, $\sim2$ Myr. This equilibration timescale is several orders of magnitude greater than the age of the nebula and indicates that explanation 3 is unlikely. 

It is difficult to rule out explanations 1 or 2. The peaks in the dust temperature at regions in the nebula consistent with outflows 2 and 3 ($\sim130$ K; Fig.~\ref{fig:PNCTOD}) are suggestive of a bipolar UV-radiation field from the central star, which favors explanation 2; however, there is no obvious enhancement in the ionizing flux along outflows 2 and 3 based on previous observations of the HII region (e.g. Latter et al. 2000, Cox et al. 2002). We favor explanation 1 given the alignment of the X-ray and high-velocity Br-$\gamma$ emission (Kastner et al. 2001; Cox et al. 2002).


\section{Conclusions}

We have reported mid-IR images of the warm dust and PAH emission from the $\sim1000$ yr-old PN NGC 7027. The key contribution of the SOFIA/FORCAST observations of NGC 7027 were the spatially resolved 33.6 and 37.1 $\mu$m images that trace the peak emission of large grains and the 6.3 and 6.6 $\mu$m images that trace the 6.2 $\mu$m PAH feature in the nebula. The mid-IR morphology of the nebula closely resembles that of the ionized inner edge of the molecular envelope as traced by Paschen-$\alpha$ line emission (Fig.~\ref{fig:PNImall}A). At the NW and SE edges of the nebula there is extended and diffuse mid-IR emission that deviates from the azimuthal symmetry of the nebula and is coincident with similar extended features in molecular hydrogen and V-band images. Notably, this asymmetry is coincident with outflow 1 from the central star (Cox et al. 2002; Fig.~\ref{fig:XRAYFCIm}).

Color temperature maps derived from the 19 and 37 $\mu$m images indicate that hot dust ($T_\mathrm{d}\gtrsim100$ K) is present in the cavity. Temperatures are also slightly higher along the outflow 1 ($T_\mathrm{d}\sim100$) relative to regions equidistant from the central star where $T_\mathrm{d}\sim70$ (Fig.~\ref{fig:PNCTOD}A). The 37 $\mu$m optical depth map exhibits a similar morphology to the CO (1-0) emission (Graham et al. 1993b) tracing the dense, molecular envelope and clearly reveals optical depth minima along outflow 1 (Fig.~\ref{fig:PNCTOD}B).

Our DustEM emission model reproduced a close fit to the ISO/SWS spectrum and SOFIA/FORCAST mid-IR photometry of NGC 7027 (Fig.~\ref{fig:PNSED}). We adjusted the dust properties of three independent components to fit the emission. The dust components were amorphous carbon large grains (LGs, $a\sim1.5$ $\mu$m) and very small grains (VSGs, $a\gtrsim12\AA$), and PAHs ($3.1\AA<a<12\AA$). The best-fit revealed a total dust mass of $5.8^{+2.3}_{-2.6}\times10^{-3}$ $\mathrm{M}_\odot$, where LGs, VSGs, and PAHs compose 96.5, 2, and 1.5 $\%$ of the dust by mass, respectively. 


Analysis of the total IR luminosity ($L_\mathrm{IR}\sim5.6\times10^3$ $\mathrm{L}_\odot$) and the hard radiation field of the $\sim200,000$ K and $7.7\times10^3$ $\mathrm{L}_\odot$ central star shows that dust in the nebula cannot be heated directly by the central star. Given the free-free emission measurements from radio observations that are unobscured from the effects of local extinction, we determined that the total luminosity in trapped Lyman-$\alpha$ photons is consistent with the observed IR luminosity. These results imply that dust in the nebula is heated by the trapped Lyman-$\alpha$ photons re-radiated in the de-excitation of ionized hydrogen. 

Lastly, we claim that the spatial anti-correlation between the optical depth and the 6.2 $\mu$m PAH emission feature strength (Fig.~\ref{fig:PNLCIm}A) is evidence for recent dust processing and rapid PAH formation along outflow 1. Despite the size of the LGs in the nebula, LGs can be destroyed by outflow 1 dust to non-thermal sputtering from ``betatron acceleration" along magnetic field lines embedded in the dense, radiatively cooled post-shock regions. The size of the LGs also makes them susceptible to frequent grain-grain collisions that fracture the LGs and redistributes mass to VSGs and PAHs. Our observations demonstrate that PAHs can indeed be formed rapidly in shocks and that grain-grain collisions are likely an efficient mechanism for PAH formation. 

There are several follow-up questions that warrant further study on dust processing and PAH formation in shocks: how efficient is PAH formation in shocks via grain-grain collisions? Do PAHs survive after forming? Can the enhancements in PAH/VSG abundances be linked to the morphology, outflow history, and/or age of PNe? Interestingly, a recent mid-IR imaging study of the $\sim2500$ yr old bipolar PN M2-9 (Werner et al. 2014) revealed that equal masses of small ($a<0.1$ $\mu$m) and large ($a>1$ $\mu$m) are present in its lobes, where they suggest collisional processing has influenced the grain size distribution. Additional spatially resolved mid-IR observations of other PNe will be important to establish the relation between central star outflows and mass redistribution in dust. Obtaining mid-IR spectra will be equally as important. Although our mid-IR observations with SOFIA are able to resolve the prominent 6.2 $\mu$m PAH emission feature, there are degeneracies in the SED modeling of these regions due to the lack of spatially resolved spectral coverage. As can be seen in Fig.~\ref{fig:PNSED}, the relative PAH and VSG abundances are difficult to fit without spectral coverage between $6\lesssim \lambda \lesssim 12$ $\mu$m. The question of the survival of newly formed PAHs is an important question to address especially given recent theoretical studies that indicate PAHs should not be able to survive $\gtrsim100$ km $\mathrm{s}^{-1}$ shocks (Micelotta et al. 2010a, b). These newly formed PAHs, however, will be difficult to trace as they move further away from the central star due to the decrease in the incident flux of optical/UV photons that excite the PAH features. Future high angular resolution, ground and space-based IR observatories such as the thirty meter-class telescopes and the upcoming 6.5-m James Webb Space Telescope (JWST; expected launch date Oct 2018) will therefore be ideal platforms for exploring the formation and evolution of PAHs in various astrophysical contexts.

\emph{Acknowledgments}. We would like to thank the FORCAST team, Terry Herter, Luke Keller, Joe Adams, George Gull, Justin Schoenwald, and Chuck Henderson, the USRA Science and Mission Ops teams, and the entire SOFIA staff. We also thank Peter van Hoof for sharing his models of the emission from NGC~7027, and the anonymous referee for the helpful comments. R. L. would like to thank Jeronimo Bernard-Salas for insightful discussion on dust processing in PNe, and Jon Livingston for help with the NGC~7027 mid-IR imaging data. This work is based on observations made with the NASA/DLR Stratospheric Observatory for Infrared Astronomy (SOFIA). This work was also based in part on observations made with the NASA/ESA Hubble Space Telescope, obtained from the data archive at the Space Telescope Science Institute, and observations made by the Chandra X-ray Observatory and published previously in cited articles. This work was carried out at the Jet Propulsion Laboratory, California Institute of Technology, under a contract with the National Aeronautics and Space Administration. SOFIA science mission operations are conducted jointly by the Universities Space Research Association, Inc. (USRA), under NASA contract NAS2-97001, and the Deutsches SOFIA Institut (DSI) under DLR contract 50 OK 0901. Financial support for FORCAST was provided by NASA through award 8500-98-014 issued by USRA.

\clearpage



\end{document}